# Evaluating the Effectiveness of Artificial Intelligence in Predicting Adverse Drug Reactions among Cancer Patients: A Systematic Review and Meta-Analysis

Fatma Zahra Abdeldjouad
*Computer Systems Engineering Department*
*National Polytechnic School of Oran - Maurice Audin*
Oran, Algeria
fatma-zahra.abdeldjouad@enp-oran.dz

Menaouer Brahami
*Computer Systems Engineering Department*
*National Polytechnic School of Oran - Maurice Audin*
Oran, Algeria
menaouer.brahami@enp-oran.dz

Mohammed Sabri
*Computer Systems Engineering Department*
*National Polytechnic School of Oran - Maurice Audin*
Oran, Algeria
mohammed.sabri@enp-oran.dz

*Abstract*—Adverse drug reactions considerably impact patient outcomes and healthcare costs in cancer therapy. Using artificial intelligence to predict adverse drug reactions in real time could revolutionize oncology treatment. This study aims to assess the performance of artificial intelligence models in predicting adverse drug reactions in patients with cancer. This is the first systematic review and meta-analysis. Scopus, PubMed, IEEE Xplore, and ACM Digital Library databases were searched for studies in English, French, and Arabic from January 1, 2018, to August 20, 2023. The inclusion criteria were: (1) peer-reviewed research articles; (2) use of artificial intelligence algorithms (machine learning, deep learning, knowledge graphs); (3) study aimed to predict adverse drug reactions (cardiotoxicity, neutropenia, nephrotoxicity, hepatotoxicity); (4) study was on cancer patients. The data were extracted and evaluated by three reviewers for study quality. Of the 332 screened articles, 17 studies (5%) involving 93,248 oncology patients from 17 countries were included in the systematic review, of which ten studies synthesized the meta-analysis. A random-effects model was created to pool the sensitivity, specificity, and AUC of the included studies. The pooled results were 0.82 (95% CI:0.69, 0.9), 0.84 (95% CI:0.75, 0.9), and 0.83 (95% CI:0.77, 0.87) for sensitivity, specificity, and AUC, respectively, of ADR predictive models. Biomarkers proved their effectiveness in predicting ADRs, yet they were adopted by only half of the reviewed studies. The use of AI in cancer treatment shows great potential, with models demonstrating high specificity and sensitivity in predicting ADRs. However, standardized research and multicenter studies are needed to improve the quality of evidence. AI can enhance cancer patient care by bridging the gap between data-driven insights and clinical expertise.

*Index Terms*—Artificial Intelligence, Adverse Drug Reactions, Cancer Patients, Predictive Models, Biomarkers

## I. Introduction

Cancer is one of the leading causes of death globally and accounts for millions of deaths annually. In 2023, almost 2 million new cancer cases and 610,000 cancer deaths were projected to occur in the United States, according to the American Cancer Society [1]. The government, society, medical industry, and scientific communities have focused significantly on reducing cancer-related mortality, anticipating the rapid development of safe and effective drugs for cancer treatment [2].

The ongoing progression in research and medicine has led to a diverse range of cancer treatments, providing patients with renewed hope and improved quality of life [3], [4]. However, oncology treatments can cause harmful side effects that may require adjustments in dosage, discontinuation of therapy, or a substitute treatment [5].

According to the World Health Organization (WHO), adverse drug reactions (ADRs) are defined as "a response to a medication that is noxious and unintended, used in man to treat" [6]. ADRs may have a significant impact on patients' quality of life and increase the burden on the healthcare system [7]. While some ADRs, such as anaphylaxis in a patient after a single exposure to an antibiotic containing penicillin, are unpredictable, many are preventable with adequate planning and surveillance [8].

Artificial intelligence (AI) is increasingly becoming an integral part of the healthcare system, rapidly transforming the way medical services are delivered [9]. In cancer research, the potential of AI has attracted significant attention for problem-solving, decision-making, and pattern recognition using predictive models [10]. AI algorithms have proven to be valuable tools in identifying and categorizing ADRs present in both single and two-drug scenarios [11].

Despite several studies on ADR prediction, no comprehensive analysis has been conducted to predict ADRs in patients with cancer using AI algorithms. Hence, this study aims to assess the effectiveness of AI models in predicting ADRs in oncology patients through a systematic review and meta-analysis by addressing the following research questions:

1) Which types of cancer have been examined using AI algorithms to predict ADRs?
2) Which cancer treatment drugs have been associated with ADRs, and what nature, frequency, and severity do these ADRs have?
3) Which biomarkers have been identified and used to predict ADRs in patients with cancer?
4) What is the performance of AI algorithms in predicting ADRs in oncology patients?

## II. METHODS

### A. Search Strategy

This study is a systematic review and a meta-analysis following the review guidelines from The Preferred Reporting Items for Systematic Reviews and Meta-analysis (PRISMA) [12], [13].

We searched the following databases for relevant literature: Scopus, PubMed, IEEE Xplore, and ACM Digital Library. The initial database search was performed between May 29 and July 31, 2023, with an updated search on August 20, 2023. The keywords used were: ("cancer patients", "oncology patients"), ("deep learning", "machine learning", "artificial intelligence", "knowledge graphs"), ("adverse drug reactions", "drug sensitivity", "drug response", neutropenia, thrombocytopenia, mucositis, "peripheral neuropathy", cardiotoxicity, "renal toxicity", nephrotoxicity, "liver toxicity", chemosensitivity, pharmacovigilance) combined with Boolean operators (OR, AND) to develop a search strategy. A detailed search strategy can be found in Appendix A. The inclusion criteria were peer-reviewed research articles published after December 31, 2017, to increase the relevance and practicality of the outcomes in today's clinical practice, in scientific journals or conference proceedings, in English, French, or Arabic, related to the use of AI for ADR prediction and focus on cancer patients. Studies were excluded if they were review papers or opinion papers, did not explicitly mention the use of AI with ADR prediction, applied to populations other than patients with cancer, were not full papers, or did not use clinical patient datasets. All results were imported into Zotero, a reference management tool, to screen the records. Duplicate studies from merging database outputs were removed.

### B. Study selection, data collection, and data extraction

The searched studies were independently screened by two reviewers (FZA and MB) who performed the eligibility criteria based on the PICO Statement [14]. A third reviewer (MS) was involved to resolve any disagreement. First, the title, abstract, and keywords were reviewed. Then, the introduction and conclusion sections were used as a second filter. Next, the full text of research papers were read to determine whether they still met the inclusion criteria. For included studies, data extraction involved authors, year of publication, country, study design, sample size, mean age, female percentage, data collection period, cancer types, reported drugs, nature, frequency, and severity of ADRs, most effective AI algorithms, and biomarkers identified and used to predict ADRs in individuals with cancer.

### C. Quality of the included studies

The quality assessment of studies was independently conducted by two reviewers (FZA and MB) at every step of the AI-based prediction models (AIPM) process, utilizing guidelines and quality standards developed for integrating AIPM into standard healthcare practices [15]. This framework enables quality assessment evaluation across six phases, including (1) data preparation, (2) AIPM development, (3) AIPM validation, (4) software development, (5) AIPM impact assessment, and (6) AIPM implementation. It was established with contributions from a group of Dutch experts who belonged to diverse professions, healthcare domains, academia, and industry sectors. A checklist for the AIPM included multiple stages, such as data collection, data preprocessing, data curation, model construction, feature selection, model evaluation, performance assessment, generalizability, validation metrics, software infrastructure, standards compliance, user interface, clinical impact, practical benefits, patient outcomes, deployment, monitoring, maintenance, and user training. Each phase has different criteria, including algorithmic bias and fairness, transparency and openness, interpretability, team members, end users and stakeholders, security, and risks. The initial three phases of the AIPM are mandatory, while the remaining phases are recommended. To simplify the process and the comparison between studies, a cumulative score of 100 points was assigned and divided as follows: 20, 20, 15, 10, 20, and 15 points for each phase, respectively, based on its importance and common integration into today's medical practices. The quality of an AIPM is based on the total score, which is categorized as poor (below 50 points), moderate (between 50 and 74 points), or strong (above 74 points). Any disagreement between the reviewers was resolved by reaching consensus through discussion.

### D. Outcomes

The performance metrics of the AI prediction model were the primary outcome measurements. Due to the characteristics of the ADR prediction problem, accuracy, sensitivity (recall), specificity, and area under the curve (AUC) were utilized. Missing data were obtained by contacting the authors. Studies that asserted two or three effective AI models or predicted more than one ADR were also included. Thus, for the meta-analysis, we incorporated studies that provided sensitivity, specificity, and AUC values with their respective sample sizes.

### E. Statistical analysis and data synthesis

Our study aimed to evaluate the AI models performance in predicting ADRs among oncology patients by analyzing the pooled sensitivity, specificity, AUC, and 95% confidence intervals (CI). To ensure the reliability of our findings, we conducted a random-effect meta-analysis and used the forest plot to report the results. We also calculated the I² statistic metric to assess the heterogeneity across studies [16]. Larger values of I² indicate significant differences in sensitivity, specificity, and AUC across studies. We also examined the results of Egger's test for small study effects to identify any publication bias. Additionally, we conducted a subgroup analysis to determine how our pooled values were affected when the AI models were applied to datasets of only one cancer type. All statistical analyses were performed in R (v4.3.1).

## III. RESULTS

### A. Selection of Included Studies

Fig. 1 illustrates the PRISMA study selection process flowchart. The search revealed 238 Scopus study titles, 55 PubMed study titles, 8 IEEE Xplore study titles, and 31 ACM Digital Library study titles. After removing duplicates, 271 papers remained; 225 were excluded for not meeting the inclusion criteria. Of these, forty-six full-text studies and three other studies identified through hand searching were reviewed. Seventeen studies met all the inclusion criteria for the systematic review. The exclusion reasons were: (i) no ADR prediction (22 studies); (ii) no use of clinical patient datasets (5 studies); (iii) no focus on cancer patients (2 studies); (iv) not a full text study (2 studies); and (v) a low quality assessment score (1 study). Ten studies were included in the meta-analysis as they provided the necessary data for calculating the pooled performance of the proposed AI models, like sensitivity, specificity, and AUC.

### B. Characteristics of the 17 Included Studies

Between 1984 and 2019, a total of 93,248 cancer patients were enrolled across 17 countries, including the United States, Canada, Denmark, China, Hong Kong, Taiwan, South Korea, Egypt, and Qatar. Out of 17 studies, 10 (59%) were published between 2022 and 2023 and covered almost all cancer types studied in the last five years (Table I). The majority of these studies (53%) were retrospective cohort studies, followed by 18% case-control studies, 12% prospective cohort studies, and 6% cross-sectional, randomized controlled, and observational studies. Electronic health records (EHR) and patient samples were utilized as data sources. The study population was predominantly female, representing 53% of the total, and had a mean age of 45 years. The quality scores of the studies ranged from 50 to 80, with an average of 64/100, indicating moderate quality of the literature (see supplementary materials in Appendix B).

### C. Cancer types studied related to ADRs prediction

Breast cancer has received the highest level of attention in relation to predicting ADRs using AI algorithms [17]–[23]. Additionally, non-small cell lung cancer (NSCLC), including lung cancer, has been the subject of extensive research in this context [17]–[19], [23]–[25]. Hematological cancers, particularly B-cell acute lymphoblastic leukemia (B-ALL) and acute lymphoid leukemia (ALL), have been investigated in-depth in [18], [19], [26]–[28]. A Danish research group has investigated testicular cancer in this context with published studies [29], [30]. Furthermore, two separate research teams have studied skin cancer, including melanoma [18], [23]. However, other cancer types, such as nasopharyngeal carcinoma (NPC) [31], kidney cancer [18], colorectal cancer (CRC) [32], pancreatic cancer [19], gastrointestinal stromal tumors (GIST) [19], and prostate cancer [23], have received relatively little attention regarding ADR prediction using AI algorithms.

### D. Cancer treatment drugs associated with ADRs

Over the last five years, researchers have utilized AI algorithms to study various anticancer drugs and predict potential ADRs. Of these drugs, the subset of cytotoxic drugs commonly used in chemotherapy, such as platinum drugs (cisplatin, carboplatin), fluoropyrimidine, anthracycline, taxanes, cyclophosphamide, 5-FU, methotrexate (MTX), and paclitaxel, have been extensively investigated in almost 76% of studies [17], [18], [20]–[24], [27]–[30], [32], [33]. Targeted therapy drugs, which include TKIs and trastuzumab (Herceptin), were studied in 23% of studies [18]–[20], [33], while only two studies have looked at immune checkpoint inhibitors, such as durvalumab and tremelimumab [18], [28]. Furthermore, only one study has examined stomach acid-related medications, including H2 blockers and PPIs, alongside other drugs that may affect CYP3A4 inducers and inhibitors [19].

### E. ADRs nature, frequency and severity

Cardiotoxicity is the second most common cause of death after a cancer diagnosis [18]. It encompasses a variety of conditions, including ischemic heart diseases (IHD), heart failure (HF), cardiomyopathy, arrhythmia, acute myocarditis, stroke, cardiogenic shock, sudden cardiac arrest, cancer therapy-related cardiac dysfunction (CTRCD), and symptomatic heart failure with reduced ejection fraction (HFrEF). Among cancer patients, cardiotoxicity was the most frequently reported ADR and was studied in 35% of the studies [18], [20], [23], [28], [32], [33]. Chemotherapy-induced neutropenia, including febrile neutropenia (FN), was the second most studied ADR, appearing in 23% of the studies [17], [22], [26], [27] and affecting 30% of the enrolled patients with varying degrees of severity. Nephrotoxicity, a severe ADR associated with cisplatin chemotherapy, was the third most studied ADR, appearing in 12% of the studies [24], [30] and affecting 20% of the enrolled patients. Sepsis, hearing loss, acute oral mucositis (AOM), hepatotoxicity, immune-mediated adverse events (imAEs), fever, and paclitaxel-induced peripheral neuropathy (PIPN) were less commonly addressed in studies [19], [21], [25]–[27], [29], [31], with varying degrees of severity.

### F. Biomarkers used in predicting ADRs

Nearly half of the studies (47%) have examined the use of biomarkers in predicting ADRs in cancer patients. Among the various biomarkers studied, ABC transporters involved in drug transport and metabolism, including ABCB1, ABCC1, ABCC2, ABCC3, ABCC4, ABCG2, ABCA10, and ABCA12, were highlighted in three out of 17 studies (18%) [21], [27], [29]. In two out of 17 studies (12%) [27], [30], genes related to drug metabolism and response, such as DHFR, MTHFR, TYMS, TMPT, NUDT15, NAT1, NAT2, and the Solute Carrier (SLC) family, which is involved in drug and nutrient transport, such as SLC16A7, SLC19A1, SLC22A11, SLCO1A2, SLCO1B1, and SLC22A2, were studied. Additionally, three out of 17 studies (18%) [28]–[30] examined other genomic markers such as SOD2, MGST3, MCM8, CNTN6, CNTN4, PI3KR2, and ZNF827. Only one study reported liver enzymes, such as AST (Aspartate Aminotransferase) and ALT (Alanine Aminotransferase) [19], and one out of 17 studies (6%) examined DNA repair genes, such as ERCC1 and ERCC2 [30]. Physiological measures such as left ventricular ejection fraction (LVEF) were studied in two out of 17 studies (12%) [22], [33].

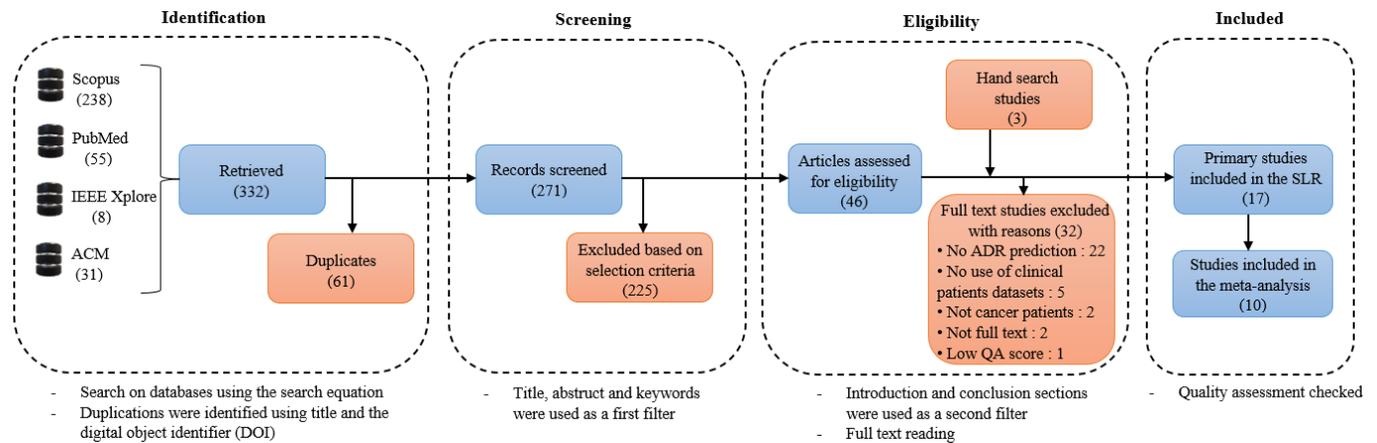

Fig. 1. Systematic literature review methodology adapted from the PRISMA principles.

### G. Meta-analysis results

The meta-analysis examined ten studies on ADR prediction, which yielded 12 performance metrics values based on different sensitivity, specificity, and AUC values for each ADR. The accuracy was not provided by the majority of studies, nor was the confusion matrix of true positive, true negative, false positive, and false negative values or the prevalence reported. Fig. 2 depicts the pooled sensitivity values, which showed a high degree of heterogeneity across studies (I² 98.2%) and a pooled sensitivity score of 0.82 (95% CI: 0.69, 0.9) from the random-effects meta-analysis. To test for small-study effects, we conducted an Egger's test and found no significant bias (p = 0.13). We excluded two studies [23], [28] that used datasets of different cancer types, as determined beforehand by the researchers. The pooled sensitivity increased to 0.84 (95% CI: 0.7, 0.92) when we excluded these studies, resulting in decreased heterogeneity (I² 94.4%) and publication bias (p = 0.10) (see supplementary materials in Appendix C).

The pooled specificity was calculated from the same studies, which showed a score of 0.84 (95% CI: 0.75, 0.9) from the random-effects meta-analysis (Fig. 3), with a high degree of heterogeneity across studies (I² 95.7%). However, we found no significant bias (p = 0.7) based on Egger's test result. By retaining only the studies that used datasets of one cancer type, we observed a slight decrease in the pooled specificity (0.83, 95% CI: 0.72, 0.91) and heterogeneity (I² 95%) but a significant reduction in publication bias (p = 0.32) (see supplementary materials in Appendix D).

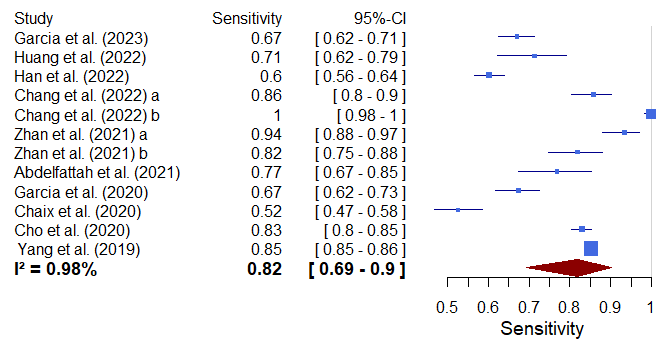

Fig. 2. Pooled Sensitivity Values from the Random-Effects Meta-Analysis of ADR Prediction Studies.

Similarly, the same studies were used to determine the pooled AUC, which showed a score of 0.83 (95% CI: 0.77, 0.87) from the random-effects meta-analysis (Fig. 4), with a high degree of heterogeneity across studies (I² 98%). We found a significant bias (p = 0.03 < 0.05) based on the Egger's test result. By retaining only the studies that used datasets of one cancer type, we observed a decrease in the pooled AUC (0.82, 95% CI: 0.76, 0.87) and heterogeneity (I² 93.7%) but no significant publication bias (p = 0.27) (see supplementary materials in Appendix E).

The proposed AI models effectively predicted ADRs among cancer patients, with sensitivity, specificity, and AUC ranging from 0.82 to 0.84.

TABLE I
SUMMARY OF INCLUDED STUDIES

| Authors (Year) | Study Design | Sample Size | Cancer Types | Reported Drugs | ADRs Nature | Effective AI Algorithms |
|---|---|---|---|---|---|---|
| Padmanabhan et al. (2023) | Retrospective Cohort Study | 513 | Hematological cancer | NR | Myelosuppression, multiple infections, febrile neutropenia, sepsis, mortality due to treatment, and infection by multidrug-resistant organisms. | XGBoost |
| Kim et al. (2023) | Case-Control Study | 10717 | Breast and lung cancer | Cytotoxic drugs | Chemotherapy-induced neutropenia | Bi-LSTM and RETAIN |
| Garcia et al. (2023) | Retrospective Cohort Study | 433 | Testicular Cancer | Cisplatin | Hearing loss | LR with Cross-Validated GWAS |
| Dong et al. (2023) | Cross-Sectional Study | 242 | Nasopharyngeal carcinoma | NR | Acute oral mucositis | GNB |
| Al-Droubi et al. (2023) | Retrospective Cohort Study | 20023 | Breast cancer, kidney cancer, B-cell lymphoma, melanoma, lung cancer | 30 drugs | Cardiotoxicity | ANN |
| Li et al. (2022) | Retrospective Cohort Study | 36030 | Colorectal cancer | Fluoropyrimidine | Cardiotoxicity | XGBoost |
| Huang et al. (2022) | Retrospective Cohort Study | 118 | Non-small cell lung cancer (NSCLC) | Platinum chemotherapy drugs | Platinum-induced nephrotoxicity | ANN-I |
| Han et al. (2022) | Retrospective Cohort Study | 703 | NSCLC, pancreatic cancer, ALL, CML, GIST, metastatic breast cancer, and other malignancies. | TKIs, anticancer drugs, H2 blockers, PPIs, other drugs and inhibitors | Hepatotoxicity | Elastic net |
| Dey et al. (2022) | Randomized Controlled Trials | 617 | Metastatic NSCLC | Durvalumab and tremelimumab | Immune-mediated adverse events (imAEs) | RF |
| Chang et al. (2022) | Prospective Cohort Study | 211 | Breast cancer | Anthracyclines, Trastuzumab, Taxanes, Cyclophosphamide, 5-FU | CTRCD and HFrEF | multilayer perceptron (MLP) model |
| Zhan et al. (2021) | Retrospective Cohort Study | 139 | B-cell acute lymphoblastic leukemia | Methotrexate | Neutropenia and fever | RF-ADASYN |
| Abdelfattah et al. (2021) | Prospective Cohort Study | 92 | Breast Cancer | Paclitaxel | Paclitaxel-induced peripheral neuropathy | The additive logistic regression model |
| Zhou et al. (2020) | Retrospective Cohort Study | 4309 | NR | Anthracycline drugs, Cyclophosphamide, Trastuzumab | Cancer therapy–related cardiac dysfunction | KNN, LR, SVM, RF, GB |
| Garcia et al. (2020) | Retrospective Cohort Study | 433 | Testicular cancer | Bleomycin-Etoposide-Cisplatin (BEP) | Nephrotoxicity | RF |
| Chaix et al. (2020) | Case-Control Study | 289 | Leukemia, Sarcoma, Neuroblastoma, Hepatoblastoma, Lymphoma, Wilms tumor | Anthracyclines | Cardiotoxicity | RF |
| Cho et al. (2020) | Observational Study | 933 | Breast cancer | Taxane-based regimen | Febrile neutropenia | XGboosting |
| Yang et al. (2019) | Case-Control Study | 17446 | Skin cancer, female breast cancer, prostate cancer, and lung cancer | Abraxane with the ingredient Paclitaxel and other drugs | Cardiotoxicity | GB |

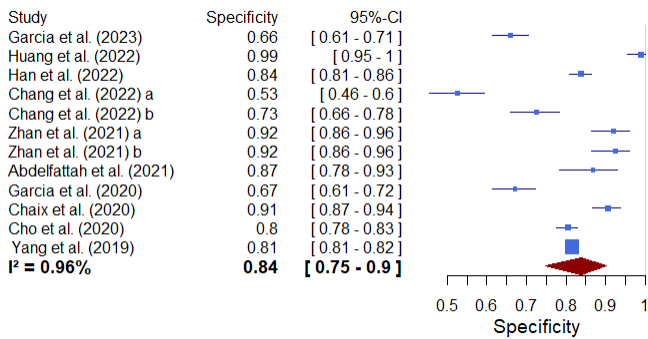

Fig. 3. Pooled Specificity Values from the Random-Effects Meta-Analysis of ADR Prediction Studies.

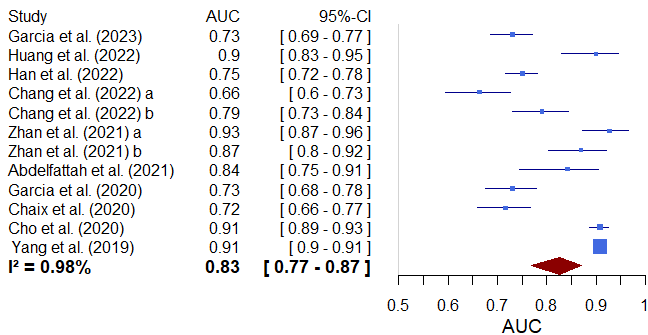

Fig. 4. Pooled AUC Values from the Random-Effects Meta-Analysis of ADR Prediction Studies.

## IV. Discussion

We report the first comprehensive systematic review and meta-analysis assessing the performance of AI models in predicting ADRs in cancer patients. This systematic review and meta-analysis indicates that AI algorithms can complement traditional clinical decisions in predicting ADRs among patients undergoing cancer treatment.

Our systematic review yielded frequent reports of cardiotoxicity, which is a life-threatening side effect, in oncology patients. As a cytotoxic drug, anthracycline is considered the primary agent responsible for cardiotoxicity induced by chemotherapy. This finding cannot be underestimated, given that anthracyclines have been among the most widely used chemotherapeutic drugs since 2012 and continue to serve as the foundation of treatment for multiple solid tumors and hematological malignancies [34]. Furthermore, breast cancer is the most common tumor examined in the last five years due to its wide prevalence among women. However, anthracyclines and taxanes are the two main drug classes used to treat breast cancer [35].

Although neutropenia and nephrotoxicity are less common than cardiotoxicity, they are no less significant. Chemotherapy induced neutropenia is a serious and common chemotherapy complication. The presence of severe neutropenia may lead to the development of fever, also known as febrile neutropenia (FN), which often requires hospitalization and the use of empiric broad-spectrum antibiotics [36]. Owing to the widespread occurrence of hematological cancer, which follows lung cancer, solid tumor malignancies have the potential to induce neutropenia by infiltrating the bone marrow. Additionally, certain lymphoproliferative malignancies, including natural killer cell lymphoma (large granular lymphocytic leukemia), hairy cell leukemia, and chronic lymphocytic leukemia (CLL), may also lead to neutropenia [37]. Nephrotoxicity is also a common ADR associated with cisplatin. Given that there is no specific treatment for renal dysfunction or injury caused by cisplatin [38], patients are typically treated with hydration and electrolyte replacement.

Although the most prevalent side effects have received the most attention, hepatotoxicity, hearing loss, and paclitaxel-induced peripheral neuropathy remain significant. Hepatotoxicity ranges from asymptomatic liver enzyme elevation to severe hepatitis [39]. Due to the rarity of these events, they may be missed in clinical settings, which is an issue that must be addressed in future studies.

ABC transporter biomarkers have been widely adopted in predicting ADRs. The 48 ABC genes in humans are organized into seven distinct families. Among these genes, 44 (from five distinct families) encode membrane transporters, several of which play a role in drug resistance and disease pathways caused by transporter dysfunction [40].

The landscape of these biomarkers is vast and diverse. Genomic markers, such as PI3KR2 and ZNF827, provide insight into an individual's genetic susceptibility to drug responses. However, their ability to predict ADRs must be validated in vivo because they have only been demonstrated to protect cardiomyocytes from cardiotoxicity in vitro by inhibiting their expression [28].

While genomic and enzyme-based biomarkers offer molecular insights, physiological measures, such as the Left Ventricular Ejection Fraction (LVEF), provide a macroscopic perspective. Evaluation of LVEF by radionuclide ventriculography (MUGA) or echocardiography is the most common method for assessing the potential cardiotoxicity of anthracyclines [41]. A decline in LVEF could be an early indication of cardiotoxic effects, allowing timely intervention to prevent irreversible cardiac damage.

The meta-analysis findings revealed the efficiency of AI models using measures such as sensitivity, specificity, and AUC. The sensitivity values varied from 0.82 to 0.84, demonstrating that the models correctly identified patients who were at risk of ADRs. Specificity values within a similar range also confirmed the models' ability to properly categorize people who would not have ADRs. AUC, a comprehensive statistic for measuring model performance, consistently showed a value of about 0.83. This outcome enhances the model's reliability across a wide range of thresholds.

However, this meta-analysis highlighted high heterogeneity across the included studies. The occurrence of heterogeneity in this context might be related to differences in the study designs, patient demographics, and methodologies employed to assess ADRs. For example, the populations studied may have a variety of genetic origins, medical histories, or cancer treatment regimens. The effectiveness of AI models is determined by the quality of data used to train them.

Heterogeneity in the assessment methods, ranging from the tools utilized to evaluate ADRs to the evaluation metrics, could also contribute to inconsistencies in the results. This variability highlights the complexity of predicting ADRs in cancer patients due to many factors. Although heterogeneity provides a comprehensive view, it complicates the interpretation of pooled results. Therefore, the findings should be interpreted cautiously, and the differences across studies should be recognized.

Excluding studies that used datasets covering various cancer types aims to minimize heterogeneity. Although this decision may strengthen the consistency of the meta-analysis, it raises concerns about the applicability of the results. Cancer is not a single, unified disease but rather a diverse group of conditions, each with its unique features.

The potential for publication bias occurs when research with positive or statistically significant findings is more likely to be published, presenting a more optimistic image than the actual situation. Although Egger's test provides some comfort, caution must be exercised when interpreting the results due to the uncertainties involved.

## V. LIMITATIONS

A careful examination of this study revealed some limitations. First, the small sample sizes in certain studies [21], [24] may create biases and impair the generalizability of the findings. Larger datasets are widely acknowledged to produce more robust and reliable results, particularly in domains such as oncology, where patient variability is crucial. Furthermore, biases, whether inherent in the research design or introduced during data collection and processing, might have affected the findings, possibly leading to overestimation or underestimation of ADRs. Finally, the employment of different prediction models across studies complicates the challenge of generating consistent findings. Although heterogeneity provides a variety of perspectives, it also makes reaching a consensus difficult.

## VI. FUTURE DIRECTIONS

Cancer constantly evolves, and our knowledge of ADRs continually expands. As such, less-studied ADRs that may be rare but have significant clinical effects require further investigation. To improve the accuracy of results, new prediction models that combine multiple types of data are needed. Multicenter studies can create larger, more diverse patient groups, reducing bias and increasing the reliability of results. The strengths and weaknesses of predictive models can be evaluated through a meta-analysis. Moving forward, integration of these models into hospital processes for real-time clinical decision-making is crucial. Additionally, educating patients on ADRs and how to manage them can empower them to play a more active role in their treatment decisions.

## VII. CONCLUSION

Although oncology treatments hold promise in fighting cancer, they are also associated with the risk of ADRs. This systematic review and meta-analysis emphasizes the significance of these reactions, with cardiotoxicity being a major concern. The search for biomarkers to predict ADRs is a promising approach for tailoring cancer treatment more effectively and reducing the risk of severe complications. Moreover, AI models can predict ADRs with high sensitivity and specificity. The use of AI in oncology has the potential to revolutionize patient care by combining human clinical decisions with machine-learned insights. However, there is a need for more standardized research in this area, given the variability across studies, to ensure the accuracy and robustness of these predictive tools. Future research should focus on multicenter studies combining multiple data types and examining the impact of AI predictive models in real-world clinical scenarios.

## REFERENCES

[1] R. L. Siegel, K. D. Miller, N. S. Wagle, and A. Jemal, "Cancer statistics, 2023," *Ca Cancer J Clin*, vol. 73, no. 1, pp. 17–48, 2023.
[2] W. Cui, A. Aouidate, S. Wang, Q. Yu, Y. Li, and S. Yuan, "Discovering anti-cancer drugs via computational methods," *Frontiers in pharmacology*, vol. 11, p. 733, 2020.
[3] A. Vitale, M. Peck-Radosavljevic, E. G. Giannini, E. Vibert, W. Sieghart, S. Van Poucke, and T. M. Pawlik, "Personalized treatment of patients with very early hepatocellular carcinoma," *Journal of hepatology*, vol. 66, no. 2, pp. 412–423, 2017.
[4] Z. Wen, S. Wang, D. M. Yang, Y. Xie, M. Chen, J. Bishop, and G. Xiao, "Deep learning in digital pathology for personalized treatment plans of cancer patients," in *Seminars in Diagnostic Pathology*, vol. 40, no. 2. Elsevier, 2023, pp. 109–119.
[5] V. Junet, P. Matos-Filipe, J. M. García-Illarramendi, E. Ramírez, B. Oliva, J. Farrés, X. Daura, J. M. Mas, and R. Morales, "A decision support system based on artificial intelligence and systems biology for the simulation of pancreatic cancer patient status," *CPT: Pharmacometrics & Systems Pharmacology*, 2023.
[6] M. A. Hadi, C. F. Neoh, R. M. Zin, M. E. Elrggal, and E. Cheema, "Pharmacovigilance: pharmacists' perspective on spontaneous adverse drug reaction reporting," *Integrated Pharmacy Research and Practice*, pp. 91–98, 2017.
[7] H. Khalil and C. Huang, "Adverse drug reactions in primary care: a scoping review," *BMC health services research*, vol. 20, no. 1, pp. 1–13, 2020.
[8] J. J. Coleman and S. K. Pontefract, "Adverse drug reactions," *Clinical Medicine*, vol. 16, no. 5, p. 481, 2016.
[9] S. Sunarti, F. F. Rahman, M. Naufal, M. Risky, K. Febriyanto, and R. Masnina, "Artificial intelligence in healthcare: opportunities and risk for future," *Gaceta Sanitaria*, vol. 35, pp. S67–S70, 2021.
[10] V. Russo, E. Lallo, A. Munnia, M. Spedicato, L. Messerini, R. D'Aurizio, E. G. Ceroni, G. Brunelli, A. Galvano, A. Russo *et al.*, "Artificial intelligence predictive models of response to cytotoxic chemotherapy alone or combined to targeted therapy for metastatic colorectal cancer patients: A systematic review and meta-analysis," *Cancers*, vol. 14, no. 16, p. 4012, 2022.
[11] S. Shankar, I. Bhandari, D. T. Okou, G. Srinivasa, and P. Athri, "Predicting adverse drug reactions of two-drug combinations using structural and transcriptomic drug representations to train an artificial neural network," *Chemical Biology & Drug Design*, vol. 97, no. 3, pp. 665–673, 2021.

[12] D. Moher, A. Liberati, J. Tetzlaff, D. G. Altman, and P. Group*, "Preferred reporting items for systematic reviews and meta-analyses: the prisma statement," *Annals of internal medicine*, vol. 151, no. 4, pp. 264–269, 2009.

[13] A. P. Siddaway, A. M. Wood, and L. V. Hedges, "How to do a systematic review: a best practice guide for conducting and reporting narrative reviews, meta-analyses, and meta-syntheses," *Annual review of psychology*, vol. 70, pp. 747–770, 2019.

[14] J. Thomas, D. Kneale, J. E. McKenzie, S. E. Brennan, and S. Bhaumik, "Determining the scope of the review and the questions it will address," *Cochrane handbook for systematic reviews of interventions*, pp. 13–31, 2019.

[15] A. A. de Hond, A. M. Leeuwenberg, L. Hooft, I. M. Kant, S. W. Nijman, H. J. van Os, J. J. Aardoom, T. P. Debray, E. Schuit, M. van Smeden *et al.*, "Guidelines and quality criteria for artificial intelligence-based prediction models in healthcare: a scoping review," *NPJ digital medicine*, vol. 5, no. 1, p. 2, 2022.

[16] J. P. Higgins, S. G. Thompson, J. J. Deeks, and D. G. Altman, "Measuring inconsistency in meta-analyses," *Bmj*, vol. 327, no. 7414, pp. 557–560, 2003.

[17] Y. Kim, Y. Lee, H. W. Park, H. Jung, Y. Hwangbo, and H. S. Cha, "Prediction of chemotherapy-induced neutropenia using machine learning in cancer patients," in *2023 IEEE International Conference on Big Data and Smart Computing (BigComp)*. IEEE, 2023, pp. 136–139.

[18] S. S. Al-Droubi, E. Jahangir, K. M. Kochendorfer, M. Krive, M. Laufer-Perl, D. Gilon, T. M. Okwuosa, C. P. Gans, J. H. Arnold, S. T. Bhaskar *et al.*, "Artificial intelligence modelling to assess the risk of cardiovascular disease in oncology patients," *European Heart Journal-Digital Health*, p. ztad031, 2023.

[19] J. M. Han, J. Yee, S. Cho, M. K. Kim, J. Y. Moon, D. Jung, J. S. Kim, and H. S. Gwak, "A risk scoring system utilizing machine learning methods for hepatotoxicity prediction one year after the initiation of tyrosine kinase inhibitors," *Frontiers in Oncology*, vol. 12, p. 790343, 2022.

[20] W.-T. Chang, C.-F. Liu, Y.-H. Feng, C.-T. Liao, J.-J. Wang, Z.-C. Chen, H.-C. Lee, and J.-Y. Shih, "An artificial intelligence approach for predicting cardiotoxicity in breast cancer patients receiving anthracycline," *Archives of Toxicology*, vol. 96, no. 10, pp. 2731–2737, 2022.

[21] N. M. Abdelfattah, M. H. Solayman, Y. Elnahass, and N. A. Sabri, "Abcb1 single nucleotide polymorphism genotypes as predictors of paclitaxel-induced peripheral neuropathy in breast cancer," *Genetic Testing and Molecular Biomarkers*, vol. 25, no. 7, pp. 471–477, 2021.

[22] B.-J. Cho, K. M. Kim, S.-E. Bilegsaikhan, and Y. J. Suh, "Machine learning improves the prediction of febrile neutropenia in korean inpatients undergoing chemotherapy for breast cancer," *Scientific reports*, vol. 10, no. 1, p. 14803, 2020.

[23] X. Yang, Y. Gong, N. Waheed, K. March, J. Bian, W. R. Hogan, and Y. Wu, "Identifying cancer patients at risk for heart failure using machine learning methods," in *AMIA Annual Symposium Proceedings*, vol. 2019. American Medical Informatics Association, 2019, p. 933.

[24] S.-H. Huang, C.-Y. Chu, Y.-C. Hsu, S.-Y. Wang, L.-N. Kuo, K.-J. Bai, M.-C. Yu, J.-H. Chang, E. H. Liu, and H.-Y. Chen, "How platinum-induced nephrotoxicity occurs? machine learning prediction in non-small cell lung cancer patients," *Computer Methods and Programs in Biomedicine*, vol. 221, p. 106839, 2022.

[25] A. Dey, M. Austin, H. M. Kluger, N. Trunova, H. Mann, N. Shire, C. Morgan, D. Zhou, and G. M. Mugundu, "Association between immune-mediated adverse events and efficacy in metastatic non-small-cell lung cancer patients treated with durvalumab and tremelimumab," *Frontiers in Immunology*, vol. 13, p. 1026964, 2022.

[26] R. Padmanabhan, A. Elomri, R. Y. Taha, H. El Omri, H. Elsabah, and A. El Omri, "Prediction of multiple clinical complications in cancer patients to ensure hospital preparedness and improved cancer care," *International Journal of Environmental Research and Public Health*, vol. 20, no. 1, p. 526, 2022.

[27] M. Zhan, Z.-b. Chen, C.-c. Ding, Q. Qu, G.-q. Wang, S. Liu, and F.-q. Wen, "Machine learning to predict high-dose methotrexate-related neutropenia and fever in children with b-cell acute lymphoblastic leukemia," *Leukemia & Lymphoma*, vol. 62, no. 10, pp. 2502–2513, 2021.

[28] M.-A. Chaix, N. Parmar, C. Kinnear, M. Lafreniere-Roula, O. Akinrinade, R. Yao, A. Miron, E. Lam, G. Meng, A. Christie *et al.*, "Machine learning identifies clinical and genetic factors associated with anthracycline cardiotoxicity in pediatric cancer survivors," *Cardio Oncology*, vol. 2, no. 5, pp. 690–706, 2020.

[29] S. L. Garcia, J. Lauritsen, B. K. Christiansen, I. F. Hansen, M. Bandak, M. D. Dalgaard, G. Daugaard, and R. Gupta, "Predicting hearing loss in testicular cancer patients after cisplatin-based chemotherapy," *Cancers*, vol. 15, no. 15, p. 3923, 2023.

[30] S. L. Garcia, J. Lauritsen, Z. Zhang, M. Bandak, M. D. Dalgaard, R. L. Nielsen, G. Daugaard, and R. Gupta, "Prediction of nephrotoxicity associated with cisplatin-based chemotherapy in testicular cancer patients," *JNCI Cancer Spectrum*, vol. 4, no. 3, p. pkaa032, 2020.

[31] Y. Dong, J. Zhang, S. Lam, X. Zhang, A. Liu, X. Teng, X. Han, J. Cao, H. Li, F. K. Lee *et al.*, "Multimodal data integration to predict severe acute oral mucositis of nasopharyngeal carcinoma patients following radiation therapy," *Cancers*, vol. 15, no. 7, p. 2032, 2023.

[32] C. Li, L. Chen, C. Chou, S. Ngorsuraches, and J. Qian, "Using machine learning approaches to predict short-term risk of cardiotoxicity among patients with colorectal cancer after starting fluoropyrimidine-based chemotherapy," *Cardiovascular toxicology*, pp. 1–11, 2021.

[33] Y. Zhou, Y. Hou, M. Hussain, S.-A. Brown, T. Budd, W. W. Tang, J. Abraham, B. Xu, C. Shah, R. Moudgil *et al.*, "Machine learning–based risk assessment for cancer therapy–related cardiac dysfunction in 4300 longitudinal oncology patients," *Journal of the American Heart Association*, vol. 9, no. 23, p. e019628, 2020.

[34] P. A. Henriksen, "Anthracycline cardiotoxicity: an update on mechanisms, monitoring and prevention," *Heart*, vol. 104, no. 12, pp. 971–977, 2018.

[35] F. Cai, M. A. F. Luis, X. Lin, M. Wang, L. Cai, C. Cen, and E. Biskup, "Anthracycline-induced cardiotoxicity in the chemotherapy treatment of breast cancer: Preventive strategies and treatment," *Molecular and clinical oncology*, vol. 11, no. 1, pp. 15–23, 2019.

[36] G. H. Lyman, E. Abella, and R. Pettengell, "Risk factors for febrile neutropenia among patients with cancer receiving chemotherapy: a systematic review," *Critical reviews in oncology/hematology*, vol. 90, no. 3, pp. 190–199, 2014.

[37] M. B. Lustberg, "Management of neutropenia in cancer patients," *Clinical advances in hematology & oncology: H&O*, vol. 10, no. 12, p. 825, 2012.

[38] X. Yao, K. Panichpisal, N. Kurtzman, and K. Nugent, "Cisplatin nephrotoxicity: a review," *The American journal of the medical sciences*, vol. 334, no. 2, pp. 115–124, 2007.

[39] M. C. Pope, M. C. Olson, K. T. Flicek, N. J. Patel, C. W. Bolan, C. O. Menias, Z. Wang, and S. K. Venkatesh, "Chemotherapy-associated liver morphological changes in hepatic metastases (calmchem)," *Diagnostic and Interventional Radiology (Ankara, Turkey)*, 2023.

[40] A. Alam and K. P. Locher, "Structure and mechanism of human abc transporters," *Annual Review of Biophysics*, vol. 52, pp. 275–300, 2023.

[41] M. Feola, O. Garrone, M. Occelli, A. Francini, A. Biggi, G. Visconti, F. Albrile, M. Bobbio, and M. Merlano, "Cardiotoxicity after anthracycline chemotherapy in breast carcinoma: effects on left ventricular ejection fraction, troponin i and brain natriuretic peptide," *International journal of cardiology*, vol. 148, no. 2, pp. 194–198, 2011.